\documentstyle[preprint,aps]{revtex}

\begin{document}

\begin{flushright}
  {\bf BU/HEPP/96-01} \\
  June 1996 \\
  hep-lat/9502015(revised)
\end{flushright}
\vspace{1.cm}
\begin{center}

{\bf {\LARGE Lattice Charge Overlap I: Elastic Limit of Pi and Rho
Mesons}}


{\bf William Andersen} \\

{\it Department of Physics, Eastern New Mexico University, Portales, NM 88130 } \\
[1em] {\bf Walter Wilcox} \\
{\it Department of Physics, Baylor University, Waco, TX 76798-7316}

\end{center}

\begin{abstract}

Using lattice QCD on a $16^{3}\times 24$ lattice at $\beta=6.0$, we examine the
elastic limit of charge overlap functions in the quenched approximation for the
pion and rho meson; results are compared to previous direct current insertion
calculations. A good signal is seen for the pion, but the electric and magnetic
rho meson results are considerably noisier. We find that the pion and rho results
are characterized by a monopole mass to rho mass ratio of $0.97(8)$ and $0.73(10)$,
respectively. Assuming the functional form of the electric and magnetic form
factors are the same, we also find a rho meson g-factor of $g=2.25(34)$,
consistent with the nonrelativistic quark model.

\vskip\baselineskip
\end{abstract}

\vfill

\section{Introduction}

The spacelike pion form factor has been the subject of a number of
completed lattice studies\cite{pion,mart1,terry1}. However, 
it has not yet been studied systematically by the method of Ref.\cite{me1}. It
would be helpful to compare the systematic and statistical characteristics of
current overlap to previous results as well as to learn about possible
limitations. Since the same set of quark propagators can also be used to construct
current overlap functions for higher spin, we investigate the rho meson as
well. As was the case with the decouplet baryons studied in Ref.~\cite{lein}, the
rho meson is not stable under the strong interactions. However, at quark masses
available in lattice calculations, it's decay is kinematically forbidden, so in
this sense our quenched results are physical. Another motivating factor in
studying the rho is to have a more \lq\lq typical" hadron to compare our pion
results to, especially in terms of the relative statistical signals. Our results
can also be used to compare to hadronic models. 

Current overlap
techniques are versatile in that the same set of quark propagators can be used to
study both elastic and inelastic processes. In the second part of this series,
referred to as Part II, the inelastic part of the correlation functions calculated
here will be used to partially address the question of the polarizability of the
charged pion. Thus, another purpose of the present study is to set the elastic
\lq\lq baseline" necessary to extract additional nonelastic properties of hadrons.
In addition to polarizability, structure functions may also be extracted using
these techniques\cite{me1,meagain}.

In summary, the purpose of this study and others like it is both as
a testing ground for lattice techniques and as a preliminary contact between
experiment and fundamental theory. As we explore the numerics, we will be more
interested in trends in our lattice data rather than the final numbers.
Nevertheless, we do not neglect comparison wherever possible to previous lattice
and experimental results. We will partially explore the systematics
associated with longer time correlations, larger volume and finite lattice
spacing effects. However, further work treating systematic effects such as
scaling and quenching will need to be done before our numerical
results can be accepted as \lq\lq physical".

We will start with a review of the formulas used in this study and then proceed
directly to the results. We finish with a summary and some brief comments about the
possible physics underlying our results.

\section{Correlation Functions}
\subsection{Formulas}

The formulas necessary for measuring the form factors of the pion and rho meson by
the current overlap technique have been developed in Ref.~\cite{me2}. The relevant
results are the following. One can define the $u,d$ flavor charge and current
density overlap matrix elements (continuum, imaginary time formalism) for
these two particles as :
\begin{eqnarray}
P(\vec{r},t)\equiv \frac{1}{2m_{\pi}}(\pi^{+}({\bf
0})|T[-\rho^{d}({\vec r},t)\rho^{u}(0)]|\pi^{+}({\bf 0})),
\\ P(\xi ;\vec{r},t)\equiv \frac{1}{2m_{\rho}}(\rho^{+}(\xi ;{\bf
0})|T[-\rho^{d}({\vec r},t)\rho^{u}(0)]|\rho^{+}(\xi ;{\bf 0})),
\\ J_{k}(\xi ;\vec{r},t)\equiv \frac{1}{2m_{\rho}}(\rho^{+}(\xi ;{\bf
0})|T[-\rho^{d}({\vec r},t)J_{k}^{u}(0)]|\rho^{+}(\xi ;{\bf 0})).
\label{eq1}
\end{eqnarray}  
$\rho^{u,d}({\vec r},t)$ represents the charge density
operator ($J_{\mu}=(\vec{J},i\rho)$);
$\rho^{+}(\xi ;{\bf 0})$ represents the vector particle state of spin polarization
$\xi$ ($\xi=\pm, 0$ refers to spin component along the z-axis) and zero momentum. 
We define the Fourier transforms:
\begin{eqnarray}
Q({\vec q\,}^{2},t) \equiv
\int d^{3}r\, e^{i{\vec x}\cdot{\vec q}}P({\vec r},t),\\
Q(\xi ;{\vec q\,}^{2},t) \equiv
\int d^{3}r\, e^{i{\vec x}\cdot{\vec q}}P(\xi ;{\vec r},t),\\
K_{k}(\xi ;{\vec q\,},t) \equiv
\int d^{3}r\, e^{i{\vec x}\cdot{\vec q}}J_{k}(\xi ;{\vec r},t).
\label{eq2}
\end{eqnarray}
In the large Euclidean time limit we can design form factor measurements
from the amplitudes (\lq\lq a" is the lattice spacing)
\begin{eqnarray}
Q({\vec q\,}^{2},t)
\stackrel{t\,\gg a}{\longrightarrow}\frac{(E_{q}+m_{\pi})^{2}}
{4E_{q}m_{\pi}}F_{\pi}^{2}(q^{2})\,e^{-(E_{q}-m_{\pi})t},\label{eqn3} \\
Q(\pm ;q_{z}^{2},t)
\stackrel{t\,\gg a}{\longrightarrow}\frac{(E_{q}+m_{\rho})^{2}}
{4E_{q}m_{\rho}}G^{2}_{\rho}(\pm;q_{z}^{2})\,e^{-(E_{q}-m_{\rho})t},\\
Q(0;q_{x,y}^{2},t)
\stackrel{t\,\gg a}{\longrightarrow}\frac{(E_{q}+m_{\rho})^{2}}
{4E_{q}m_{\rho}}G^{2}_{\rho}({0;q_{x,y}^{2})\,e^{-(E_{q}-m_{\rho})t},}\\
Q(\pm ;q_{x,y}^{2},t)
\stackrel{t\,\gg a}{\longrightarrow}\frac{(E_{q}+m_{\rho})^{2}}
{4E_{q}m_{\rho}}G^{2}_{\rho}(\pm;q_{x,y}^{2})\,e^{-(E_{q}-m_{\rho})t},\\
K_{y}(+;q_{x},t)
\stackrel{t\,\gg a}{\longrightarrow}iq_{x}\frac{(E_{q}+m_{\rho})}
{4E_{q}m_{\rho}}H_{\rho}(+;q_{x}^{2})\,e^{-(E_{q}-m_{\rho})t}.
\label{eqn4}
\end{eqnarray}
$F_{\pi}$ is the pion electric form factor and one can show that
$G_{\rho}(\pm;q_{z}^{2})=G_{\rho}(0;q_{x,y}^{2})$. In Eq.(\ref{eqn4}) we may
cyclically change the directions of the current, polarization and momentum;
anticyclic changes produce a minus sign. We can then extract the charge ($G_{c}$),
quadrupole ($G_{q}$) and magnetic ($G_{m}$) form factors of the rho meson from
(the common magnitude of spatial momentum is $\bar{q}$)
\begin{eqnarray}
G_{c}=G_{\rho}(\pm;q_{z}^{2})(\frac{2}{3}+
\frac{1}{3}\eta),
\label{4.5}\\ 
G_{q}= \frac{m(E_{q}+m)}{\bar{q}^{2}}G_{\rho}(\pm;q_{z}^{2})( -1 +
\eta),
\label{eqn5}\\ G_{m}=\frac{2H_{\rho}(+;q_{x}^{2})}{G_{\rho}(\pm;q_{z}^{2})( 1 +
\eta)},
\label{eqn6}
\end{eqnarray}
where
\begin{equation}
\eta\equiv
(2\frac{G^{2}_{\rho}(\pm;q_{x,y}^{2})}{G^{2}_{\rho}(\pm;q_{z}^{2})}-1)^{1/2}.
\label{chi}
\end{equation}

\subsection{Simulation Details}

Our quenched lattices were constructed by the algorithm in Ref.~\cite{cabbibo},
thermalized by $11000$ sweeps, and separated by $1000$ sweeps. (Ten of these
lattices overlap with those used in Ref.~\cite{nucleon}.)
Our results are obtained with Wilson fermions on twenty
$16^{3}\times 24$ lattices at $\beta=6.0$. We used the exactly conserved lattice
charge and current densities in forming the overlaps. For the quarks, we used
periodic boundary conditions in space and \lq\lq fixed" boundary conditions in
time (lattice time boundary gauge field time links are unused). When a particular
polarization and momentum state is called for in the above, we measure this in all
possible ways in a given configuration in order to reinforce the signal. For
example, the $G_{\rho}(\pm;q_{z}^{2})$ amplitude can be measured by averaging the
results for $\xi =\pm$. An additional two measurements are afforded by
$G_{\rho}(0;q_{x,y}^{2})$. When a correlation function for the second momentum is
constructed, we combine signals from
$\vec{q}=(\pm\frac{\pi}{8},\pm\frac{\pi}{8},0)$,
$(\pm\frac{\pi}{8},0,\pm\frac{\pi}{8})$ and
$(0,\pm\frac{\pi}{8},\pm\frac{\pi}{8})$.

In forming these amplitudes on the lattice, we neglect the charge and current
self-contraction loops, which do not vanish in this context. In terms of quark
lines, these are the disconnected (sea quark) diagrams, which are extremely
difficult to simulate$^{1}$. However, it is unlikely that such local objects
will significantly affect the elastic (large time separation) limit studied here.
The connected amplitude and an example of a disconnected contribution are
shown in Figs.1(a) and (b), respectively. In collecting data from our
propagators, we do so in a manner which is as symmetrical as possible in
time, given that the conserved lattice charge density is nonlocal. Thus,
we use the symmetrized charge density, $\frac{1}{2}[\rho (\vec{r},t)+\rho
(\vec{r},t-1)]$, naturally associated with integer time locations, in forming the
$H_{\rho}(\xi;\vec{q}~^{2})$ correlations. In addition, we find it is important
to locate the two currents as symmetrically as possible between the particle
interpolation fields, which are fixed in time. When a symmetrical time array of
currents is not possible, we average over the two possibilities on each
configuration. The time behavior of our signals is then much smoother.
 
Our lattice interpolation fields for the pion and rho are standard,
given by $\bar{\psi}^{d}\gamma_{5}\psi^{u}$
and $\bar{\psi}^{d}\epsilon_{i}(\xi)\gamma_{i}\psi^{u}$, where
$\hat{\epsilon}(\pm)=\frac{1}{\sqrt{2}}(\mp\hat{x}-i\hat{y})$,
$\hat{\epsilon}(0)=\hat{z}$ (rho meson rest frame). However, our
interpolation fields are smeared over the entire lattice spatial volume (using the
lattice Coulomb gauge), projecting onto zero momentum. Form factors are calculated
at three values of the hopping parameter ($\kappa=0.148$, $0.152$ and $0.154$) and
parameters are extrapolated linearly to the chiral limit.

The lattice analog of the coordinate space functions $P(\vec{r},t)$,
$P(\xi ;\vec{r},t)$ and $J_{k}(\xi ;\vec{r},t)$, which we shall
denote as ${\cal P}(\vec{r},t)$, ${\cal P}(\xi ;\vec{r},t)$ and
${\cal J}_{k}(\xi ;\vec{r},t)$, can be shown to be real and purely
imaginary, respectively, in the configuration average. That is, using
the identity developed in Ref.~\cite{new2} and given that the gauge fields $U$ and
$U^{*}$ appear with equal weight in the ensemble average, one may show that
\begin{eqnarray}
{\cal P}(\vec{r},t;\{U\})={\cal P}(\vec{r},t;\{U^{*}\})^{*},
\label{eqnP} \\
{\cal P}(\xi ;\vec{r},t;\{U\})={\cal P}(\xi
;\vec{r},t;\{U^{*}\})^{*},
\end{eqnarray}
and
\begin{equation}
{\cal J}_{k}(\xi ;\vec{r},t;\{U\})=-{\cal
J}_{k}(\xi ;\vec{r},t;\{U^{*}\})^{*}.
\label{eqnJ}
\end{equation}
These identities allow us to neglect terms which are purely noise in the
discrete Fourier transforms of ${\cal P}(\vec{r},t;\{U\})$, ${\cal
P}(\xi ;\vec{r},t;\{U\})$ and ${\cal J}_{k}(\xi ;\vec{r},t;\{U\})$. (${\cal
P}(\vec{r},t)$, ${\cal P}(\xi ;\vec{r},t)$ and ${\cal J}_{k}(\xi ;\vec{r},t)$ are
clearly bad notation for these four-point functions which obviously depend on more
than just the relative spatial and time separations. However,
Eqs.(\ref{eqnP})-(\ref{eqnJ}) convey the basic idea.) Also, by
combining Fourier transform measurements involving both $\vec{q}$ and $-\vec{q}$
as we do, one is explicitly symmetrizing or antisymmetrizing these, so there is no
need to assume evenness or oddness in $\vec{r}$. We can actually show that the
identities,
\begin{eqnarray}
{\cal P}(\vec{r},t)={\cal P}(-\vec{r},t)^{*}, \\
{\cal P}(\xi ;\vec{r},t)={\cal P}(\xi ;-\vec{r},t)^{*},
\end{eqnarray}
hold exactly, configuration by configuration, for the lattice charge overlap
functions. We can find no corresponding exact statement for ${\cal
J}_{k}(\xi ;\vec{r},t)$, however.

One important aspect of forming the overlap functions is to have an efficient
algorithm for sewing together the quark lines to form the charge and current
densities at all relative spatial separations. Our analysis codes use the
following trick to save computer time. The inner loop has a sum that looks like:
\begin{equation}
f(\vec{r})= \sum_{\vec{x}}f2(\vec{x}+\vec{r})f1(\vec{x}).\label{eqnvec}
\end{equation}
($\vec{x}$, $\vec{r}$ are spatial locations in a cubic 3D space.) This loop,
which scales like $N_{s}^{2}$, where $N_{s}$ is the number of space points in the
lattice, and must be repeated at all chosen relative time separations, is not
easily vectorizable for spatially periodic boundary conditions. This equation was
replaced by a version using Fourier transforms:
\begin{equation}
f(\vec{r})= \sum_{\vec{q}}e^{-i\vec{q}\cdot\vec{r}}Q2(-\vec{q})Q1(\vec{q}),
\end{equation}
where
\begin{eqnarray}
Q1(\vec{q})=
\sum_{\vec{x}_{1}}e^{-i\vec{q}\cdot\vec{x}_{1}}f1(\vec{x}_{1}), \\
Q2(-\vec{q})= \sum_{\vec{x}_{2}}e^{i\vec{q}\cdot\vec{x}_{2}}f2(\vec{x}_{2}).
\end{eqnarray}
This version is easily vectorized (i.e., the $Q2(-\vec{q})$ times $Q1(\vec{q})$
above) and can utilize fast Fourier transform routines$^{2}$. This
technique allows us to extract, with only a small amount of computer time, the
full time extent of the current overlap correlations, even up to $t=22$, where the
currents begin to overlap with the interpolation fields.

In making our choice of fits to masses and correlation functions, we examine the
chi-squared per degree of freedom, $\chi^{2}_{d}$, using correlated fits. The
covariance matrix\cite{jack}, $C_{ij}$, is estimated from the
single-elimination jackknife\cite{efron,bernard}:
\begin{equation}
C_{ij}= \frac{N-1}{N}\sum_{n=1}^{N}(X_{i}(n)-\hat{X}_{i})
(X_{j}(n)-\hat{X}_{j}),\label{it1}
\end{equation}
where $N$ is the number of configurations, $i$ and $j$ label
different time slices, and the $X_{i}(n)$ represent jacknifed
propagator data, $\hat{X}_{i}$ being the average to remove the bias.
The $\chi^{2}$ is then given by
\begin{equation}
\chi^{2}= \sum_{i,j}C^{-1}_{ij}(\bar{Y}_{i}-f_{i})(\bar{Y}_{j}-f_{j}),\label{it2}
\end{equation}
where the $\bar{Y}_{i}$ are the average experimental values of the propagators
and the $f_{i}$ are the functional form values for these time slices.
However, it has been noticed by many authors that the covariance matrix
overestimates correlations on propagators for small numbers of
configurations\cite{michael}. Therefore our procedure on fits is the following.
We choose our time intervals based upon obtaining acceptable 
values of the $\chi^{2}_{d}$. However, following the suggestion
in Ref.~\cite{michael}, the actual fits themselves are uncorrelated,
which simply means neglecting the off-diagonal components
of the $C_{ij}$. We feel it is necessary, and will see later it
can be important, to take correlations into account in extrapolating our 
results across $\kappa$ values. Therefore, the full forms of
Eqs.(\ref{it1}), (\ref{it2}) are used in our chiral extrapolations ($i,j$ then
label $\kappa$ values).

A third order single elimination jackknife was used for error
analysis; the first order defines error bars on the time correlation
functions, the second defines error bars on the time correlation fits and the
third is necessary for the chiral extrapolation of the results. Our
masses were measured on twenty gauge configurations using single exponential fits
and are listed in Table I along with the chi-squared per degree of freedom,
$\chi^{2}_{d}$, of the fit. The source position for these propagators
is the first time slice of the lattice ($t\equiv 0$). The masses
are consistent with those measured with twelve configurations in
Ref.~\cite{nucleon}. However, in order to find a time interval on which the
correlated $\chi^{2}_{d}$ was acceptable for our pi and rho masses simultaneously,
we had to increase the time interval considerably beyond the fits in this
reference, resulting in slightly increased error bars.

Two definitions for the Wilson quark mass, which agree for
$\kappa\approx \kappa_{c}$, have been used in previous studies of
form factors. These definitions are
\begin{equation}
ma=\frac{1}{2}(\frac{1}{\kappa}-\frac{1}{\kappa_{c}}),\label{eqn7}
\end{equation}
and
\begin{equation}
ma=ln(\frac{4\kappa_{c}}{\kappa}-3),\label{eqn8}
\end{equation}
where $\kappa_{c}$ is the critical $\kappa$ value for which the pion mass
vanishes. The first definition is motivated by leading order chiral perturbation
theory; the second definition comes from tadpole improving\cite{tad} the quark
pole mass in free field theory. Fitting the square of the pion mass in Table I
with Eq.(\ref{eqn7}) gives {$\kappa_{c}=0.1566(2)$. (Ref.~\cite{bernard} gives
$\kappa_{c}=0.1570(1)\pm .0002$ from $t= 9$ to $14$ point-source fits, whereas the
fits here are $t= 15$ to $18$ smeared-source. See the next
section for more comments on our time fit choice.) Fitting the same data to
Eq.(\ref{eqn8}) gives $\kappa_{c}=0.1564(2)$. We prefer Eq.(\ref{eqn8}) and the
latter $\kappa_{c}$ value in this work to partly correct for the fact that our
quark masses may not be in the leading order chiral limit range.

As was done in Ref.~\cite{nucleon}, we shall concentrate here on
dimensionless quantities in extrapolating to the chiral limit. This is because
a ratio of similar physical quantities is often less subject to
systematic errors. Our philosophy in comparing our results to experiment is to
assume the simpliest possible functional form and to extrapolate the fit
parameters rather than individual form factor values in order to make contact with
phenomenology. This is crucial in the case of the pion, where the four momentum
transfer in the chiral limit vanishes. 

\section{Results}
\subsection{Preliminaries}

First, let us make a point about our correlation functions, which are
four-point functions. In order to measure current overlap functions, it is
necessary to form an amplitude which looks generically like Fig.1(a).
Eqs.(\ref{eqn3})-(\ref{eqn4}) tell us the form factors are identified in the large
time separation limit, $t$, of the charge or current densities. However, large time
separation between the currents produces small time separation between the fixed
interpolation fields and the currents. It is only as we increase the time
separations between all four time locations of our four-point function that the
lattice amplitudes project with increasing accuracy on the ground state. This is
quite different from generic two-point functions which, outside of time boundary
effects, are guaranteed to have a better ground state overlap
for larger time separations. This has important implications for our time fits, as
we will discuss below.

We next discuss two tests done to help motivate the choices made in this study
regarding the use of smeared fields and for the time position of the interpolation
fields.

One test is given in Fig.2(a), which shows a graph of the Fourier transform of
the equal-time charge overlap function, $Q(\vec{q\,}^{2},t)$, as both charge
densities are moved {\it in unison} between pion source and sink at
$\kappa=0.154$. This measurement is associated with half-integer time steps since
the conserved lattice current is non-local in time. We are using smeared-to-smeared
quark fields and are examining the two lowest momenta on a linear scale using
twenty configurations. The data should be flat if the lattice is long enough in
time. The satisfactory results seen in Fig.2(a) are to be contrasted with those in
Fig.2(b), which repeat the same measurement (using a subset of ten
configurations), but using smeared-to-point$^{3}$ correlation functions.

Having adopted smeared fields, in Fig.3 we show measurements of the
Fourier transform of the pion overlap distribution functions, $Q({\vec
q\,}^{2},t)$, on a ${\rm log_{10}}$ scale at the two lowest spatial momenta
transfer for $\Delta T=17$ (boxes) and $\Delta T=23$ (diamonds), again at
$\kappa=0.154$. $\Delta T$ refers to the time separation of the pion source and
sink fields; time separations between the charge densities range from 0 to
10. (The $\Delta T=17$ time correlation data is from
Ref.~\cite{me1}, where only charge overlap time separations up to $10$ were
considered because of the expensiveness of constructing the overlap function. See
Section II(B) above for comments on the improved method used here.) Note that 
the $\Delta T=17$ measurements use twelve
configurations and the $\Delta T=23$ use twenty, ten of which overlap. 
There are no significant differences within statistical errors for the lowest
momentum (upper set of points), but there appears to be a difference
larger than the error bars on the second momentum results. To examine the
situation from another point of view, consider Figs.4(a) and (b). Here
we have plotted the local $(E-m)$ determined from the exponential falloff
of the pion overlap function for the three $\kappa$ values in this study
in (a) for $|\vec{q}_{min}|=\frac{\pi}{8}$ and (b) for
$|\vec{q}|=\sqrt{2}|\vec{q}_{min}|$. This measurement is also naturally
associated with half-integer time values. Although the different $\kappa$ values
approach the asymptotic limit at different rates, all of the first momentum results
are quite consistent with the expected exponential falloff (determined from
continuum dispersion and the measured pion mass) by a time separation of
$14$ steps. The second momentum results in Fig.4(b) are more problematical.
The $\kappa=0.154$ local $(E-m)$ results are seen to eventually approach the
correct asymptotic falloff (slower than for the first momentum), albeit with large
error bars. However, the smallest $\kappa$ value at this larger momentum is
apparently displaying a violation of continuum dispersion. (The
dotted lines in Figs.4(a) and (b) give the spin $0$ lattice result; see
Eq.(\ref{spin0}) below.) Because of this {\it apparent} violation we prefer in this
study to exclude higher momentum measurements for both the pion and rho meson. (We
will estimate the remaining effect of finite lattice spacing on our first momentum
extrapolated quantities at the end of the next section.) An analysis of the
lowest momentum rho meson local electric $(E-m)$ shows a similar
pattern to the pion with, however, larger errors. In addition, we will see
later that the rho meson magnetic $(E-m)$ becomes consistent with continuum
dispersion much earlier in time than the electric ones. Thus, our lattice appears
to have enough time steps to filter out asymptotic limit of the first momentum in
the overlap correlation functions and continuum dispersion is reliable within
errors.

\subsection{Numerical Results}

After these preliminaries, we now begin with a survey of the correlation functions
measured in this study, given by the quantities in Eqs.(\ref{eqn3})-(\ref{eqn4}).
The different $\kappa$ values are combined and shown in Figs.5-8.
We will take some pains to examine all the correlation functions since this is the
first complete study using the current overlap technique.

First, let us explain our fitting procedure. As pointed out
above, large time separations of the currents move them toward the interpolation
fields, fixed at the time ends of the lattice. There is no {\it \'a priori} reason
that the large time separation amplitudes are to be favored; non ground state
contributions are expected both for small as well as for large time separations.
We will use the calculated $\chi^{2}_{d}$ of the fits to determine
allowed fit time intervals for the correlation functions. In our fits, we wish to
extend the time plateaus sufficiently that we are testing the fit in a
nontrivial way, but not so extended that the $\chi^{2}_{d}$ becomes unmanageable or
the statistical signal has decayed. Our choice of fitting $3$ time values on
overlap functions and $4$ on mass correlation functions is thus a compromise
between systematics and statistics. (Both types of fits involve $2$ independent
degrees of freedom.) Our overlap fits will assume the continuum dispersion
relation, $E^{2}={\vec{q}\,}^{2}+m^{2}$. We attempt to fit all $\kappa$ values for
given type of correlation function on the same time interval in order to increase
possible correlations across $\kappa$. We find this be be a significant
effect for the chiral extrapolation of the pion data. (See the discussion under
Fig.9 below.)

Fig.5 shows the correlation functions, $Q({\vec{q}\,}^{2},t)$ versus time
separation, $t$, for the pion. (The next $4$ figures are ${\rm log_{10}}$ plots.)
The lines drawn are the best fits over a common $t=14$ to $16$ time separation
interval. A single exponential form is evident even for large time separations. The
form factors, $F$, from these fits using Eq.(\ref{eqn3}) are listed in Table II
along with the implied $F/F^{VD}$ ratio ($F^{VD}$ is the vector dominance result
predicted by the masses in Table I) and the $\chi^{2}_{d}$. Notice that all of
the $F/F^{VD}$ ratios are low compared to the vector dominance result of $1$.

Fig.6 shows the time separation correlation function for the rho meson
corresponding to adding the signals for $Q(\pm ;q_{z}^{2},t)$ and
$Q(0;q_{x,y}^{2},t)$ (see Eqs.(8) and (9).) The error
bars are significantly larger than for the pion. The $\kappa=0.148$ and $0.152$
results are fit across $t=14$ to $16$ (as for the pion) with acceptable
$\chi^{2}_{d}$ values; see the $\chi^{2}_{d}$ values in Table IV under the line
for $G_{c}/G^{VD}$. However, a similar fit to the $\kappa=0.154$ data gives a
rather large value for $\chi^{2}_{d}$ ($1.99$) caused by poor exponential
behavior near the time separation edge. We have therefore moved this fit inward
until an acceptable fit is achieved. This occurs after shifting it a single time
step ($t=13$ to $15$ fit); we now have $\chi^{2}_{d}=0.43$. We note that the local
$(E-m)$ values for $\kappa=0.154$ are low, but consistent within large error bars
with continuum dispersion across these time slices; the other two $\kappa$ values
are also consistent in this sense.

Fig.7 shows the magnetic correlation function
$K_{y}(+;q_{x},t)$ as a function of time separation. The time behavior
of these functions, which come from charge-current overlap (see Eqs.(\ref{eq1}) 
and (\ref{eq2})), is quite different from the electric (charge-charge overlap)
functions just examined. We are able to fit these functions much earlier in time
to the required exponential behavior with acceptable $\chi^{2}_{d}$ values than in
Figs.5 or 6, resulting in improved error bars. The situation is similar to the
magnetic correlation functions for the proton, examined in Ref.~\cite{nucleon},
which also assumed their expected exponential behavior more quickly than their
electric counterparts. The $\chi^{2}_{d}$ values of these fits are listed in Table
IV (under the line for $G_{m}$). Again, the local $(E-m)$ values across the chosen
time slices are consistent within error bars with continuum dispersion.

Fig.8 shows the rho correlation function $Q(\pm ;q_{x,y}^{2},t)$ as a function
of time separation. There seems to be a nonexponential systematic effect for larger
time separations for $\kappa=0.152$ and $\kappa=0.154$. In addition, the local
$(E-m)$ plots for these $\kappa$ values are never consistent within error bars
with the expected exponential falloff. We feel that the appropriate conclusion
from these difficulties is that the quadrupole form factor can not be reliably
extracted with our data sample. We note that if the quadrupole form factor is zero,
the results in Figs.6 and 8 should be the same within statistical errors for
each $\kappa$ value; there is indeed strong overlap of the error bars. Therefore,
in what is to follow, we assume that $G_{q}(q^{2})=0$, consistent with
the data$^{4}$. In Eqs.(\ref{4.5})-(\ref{eqn6}) this makes $\eta=1$. The values for
the electric, $G_{c}$, and magnetic, $G_{m}$, form factors in Table IV were
extracted with this assumption.

Now that we have examined the correlation functions, in Fig.9 we begin to show
results of extrapolations of the pion correlation functions. The form factors from
Table II are characterized by the corresponding value of the monopole mass,
$m_{M}$, implied by ($q^{2}< 0$)
\begin{equation}
F_{\pi}(q^{2})=(1-\frac{q^{2}}{m_{M}^{2}})^{-1}.\label{monopole}
\end{equation}
The $m_{M}$ values arising from the form factors are divided by the
measured $m_{\rho}$ and displayed in Table III and Fig.9. Also shown are pion
results from Ref.~\cite{terry1}, which used $28$ gauge field configurations
of size $24^{2}\times 12^{2}$ including an extra quark propagator with reversed
momentum to decrease error bars. (Values and error bars for the pion and rho meson
are extracted from the charge radius results in Table 1 of Ref.~\cite{terry1} using
$m_{M}=\sqrt{6}/r$.) The result of our analysis is linearly extrapolated to the
chiral limit; the $\chi^{2}_{d}$ is $0.53$. Our final result for the
$m_{M}/m{\rho}$ ratio is $0.97(8)$. Note that the error bars on $m_{\rho}$ are not
included in the our results; this is not the conservative assumption. However, one
reason for this is so our error bars can be compared with those in
Ref.~\cite{terry1}. In addition,
$m_{M}$ and $m_{\rho}$ are likely to be strongly correlated in both studies, so
$m_{M}/m_{\rho}$ is probably better determined than uncorrelated addition of the
error bars would indicate. Also note that the physical $q^{2}$ of
Ref.~\cite{terry1}, set by the nucleon mass, is $|q^{2}|\approx 0.16$ GeV$^{2}$.
If we set our scale in the same fashion using the results of
Ref.~\cite{physrevlet}, our physical momentum is $|q^{2}|\approx 0.47$ GeV$^{2}$,
approximately three times larger. Ref.~\cite{terry1} uses
$\beta=5.9$ as opposed to $6.0$ here which means that the quark mass times lattice
scale values, $ma$, from Ref.~\cite{terry1} must be multiplied by the ratio of
scales so that we can compare results at the same physical quark mass value.
(Ref.~\cite{terry1} uses Eq.(\ref{eqn7}) for $ma$.) The lattice scale in
Ref.~\cite{physrevlet} is reasonably consistent within large errors with
Ref.~\cite{terry1} using asymptotic scaling. The nucleon mass times the lattice
scale, $m_{N}a$, was $0.54(3)$ in Ref.~\cite{physrevlet}, as opposed to $0.59(6)$
in Ref.~\cite{terry1}, giving the ratio of scales,
$a(\beta=5.9)/a(\beta=6.0)=1.09(13)$. Quenched asymptotic scaling would predict
$1.12$.

Let us make several additional points about the data sets shown in Fig.9. First,
since the physical momentum in these two studies are so different, it
is not required that the results be in agreement with one another at
a given $ma$ value. Second, since the earlier study did not use correlated
$\chi^{2}$ values to monitor the quality of their fits and because of the
different lattice sizes and $\beta$ values, a comparison between the
error bars can not be made directly. Ref.~\cite{terry1} does not give
chiral charge radii, but we will do a rough extrapolation of their implied
$m_{M}/m{\rho}$ values in the next section to compare with ours. Third, the
correlated fit shown in Fig.9 is quite different from what one would expect from
an uncorrelated fit of the displayed error bars. Note that this is not a
special result on the selected correlation time interval ($t=14$ to $16$), but
occurs generally in our data whenever the same $3$-time-slice interval is
selected across all $\kappa$ values.

Experimentally, a monopole form for the spacelike pion form factor
gives an excellent fit to the data with, however, a pole mass about $4\pm
1\%$ low compared to the mass of the rho\cite{experiment}. This value was arrived
at from the uncertainty in the fit pole mass given in the first paper in
Ref.~\cite{experiment}: $736\pm 9$ MeV. (The value of the $\rho^{0}$ mass in the
1994 Particle Data Group review\cite{review} is given as $768.1\pm 1.3$ MeV.) Note
the agreement in the pole mass value  in the two studies in Ref.~\cite{experiment}
despite the difference in the studied
$|q^{2}|$ ranges. Thus, it is encouraging that the chiral extrapolation of our
results agrees in direction and approximate magnitude with the small n\"aive
vector dominance violation seen in experiments. Of course, this may simply be
fortuitous since our error bars are about twice the size of the experimental pole
mass shift.

In order to compare with the pion case, we again examine a monopole mass
fit for the rho meson lowest momentum. Fig.10 shows the results and Table V
lists the $m_{M}/m_{\rho}$ ratios for the three $\kappa$ values of this study.
The linear chiral extrapolation of the data gives $0.73(10)$. The $\chi^{2}_{d}$
for the correlated fit is $0.13$. It is clear that the error bars here are
significantly larger than for Ref.~\cite{terry1}. In addition, although both
studies are low compared to vector dominance, our final extrapolation is about
$2.7\sigma$ away from the expected vector dominance result of $m_{M}/m_{\rho}=1$.
Again, note our point above regarding the physical $|q^{2}|$ values of the
two studies.

The last quantity we examine is the rho meson magnetic form
factor, values of which are presented in Table IV. We note that although the
conserved lattice current density is extended in space, the spatial separations
used in the Fourier transform (see Eq.(\ref{eqn4}) above) are unambiguous
since the Fourier transform is taken in directions perpendicular to the Lorentz
index. Unlike the electric form factor, we have no absolute normalization for the
$q^{2}=0$ value of $G_{m}(q^{2})$. To help evaluate the significance of the data,
we make the assumption that the electric and magnetic form factors have the same
$q^{2}$ dependence for our range of momentum transfer. That is, independent of
the unknown functional form, we simply assume ($g_{\rho}$ is a constant
in $q^{2}$)
\begin{equation}
\frac{G_{m}(q^{2})}{G_{c}(q^{2})}= g_{\rho},\label{geqn}
\end{equation} 
at each value of $\kappa$, and look at the dependence of $g_{\rho}$ on the quark
mass. In the nonrelativistic limit we expect that $g_{\rho} \approx 2$. The
nonrelativistic quark model determines the magnetic moment in terms of the
constituent quark mass, $m_{q}$, as
\begin{eqnarray*}
\mu_{\rho} = \frac{e}{2m_{q}}.
\end{eqnarray*}
Experimentally, $\mu_{\rho} = g_{\rho}e/2m_{\rho}$, so that $g_{\rho} =
m_{\rho}/m_{q}$, which is rendered less than $2$ by the binding energy of
the rho. See also Ref.~\cite{g} for a prediction from a light-front model, in
excess of $2$.

In Fig.11 and Table V we examine the g-factor that results. This quantity seems
to approach it's final value from below. After doing a linear extrapolation,
we find that $g_{\rho}= 2.25(34)$. The $\chi^{2}_{d}$ on this correlated fit is
$0.03$.

In order to estimate the size of the finite lattice spacing systematic errors in
this simulation, we used the lattice spin $0$ dispersion
relation\cite{new4}
\begin{equation}
sinh^{2}(\frac{Ea}{2}) = sinh^{2}(\frac{ma}{2}) +
\sum_{i}sin^{2}(\frac{p_{i}a}{2}),
\label{spin0}
\end{equation}
and the normalized lattice propgator for a particle of mass $m_{M}$
\begin{equation}
F^{L}_{\pi}(q^{2})\equiv
\left(1-\frac{sinh^{2}(\frac{Ea}{2}) -
\sum_{i}sin^{2}(\frac{p_{i}a}{2})}{sinh^{2}(\frac{m_{M}a}{2})}\right)^{-1},
\end{equation}
(the analog of Eq.(\ref{monopole})) to
recalculate the fits in the correlation functions Eqs.(\ref{eqn3})-(\ref{eqn4})
above. (The dimensionless mass parameter,
$\hat{M}$, in Eq.(3.\,17) of Ref.~\cite{new4} is related to the mass defined by the
exponential falloff of the two-point function, $\hat{m}$ by $\hat{m}=
2sinh^{-1}(\frac{\hat{M}}{2})$.) Also, we made the continuum to lattice
replacements\cite{new5},
\begin{eqnarray}
p_{i} \longrightarrow 2a^{-1}sin(\frac{p_{i}a}{2}),\\
E \longrightarrow 2a^{-1}sinh(\frac{Ea}{2}),
\end{eqnarray}
for the kinematic factors in these equations. The effect of these replacements
was to decrease the chirally extrapolated monopole to rho mass ratio,
$m_{M}/m_{\rho}$, by approximately $3.6\%$ for the pion and $2.7\%$ for
the rho meson. The extrapolated rho meson g-factor was increased by $6.6\%$ by
these modifications.

To partially investigate the effect of inserting lower momentum, we
evaluated the form factors at $\kappa=0.152$ on lattices of size
$20^{3}\times 30$. These lattices were also generated at $\beta=6.0$.
They were thermalized by $5000$ sweeps and separated by $1000$ pseudo-heatbath
sweeps. The measurements on these larger lattices were done as similar as
possible to our $16^{3}\times 24$ lattice measurements to provide a direct
comparison. Using the same scale as before, the lowest momentum on
these lattices is $|q^{2}|\approx 0.30$ GeV$^{2}$, still about twice that in
Ref.~\cite{terry1}. The dimensionless pion and rho masses measured using 50
configurations on $t=15$ to $18$ fits ($t=0$ again labels the time origin) are
given by $0.479(4)$ and $0.547(6)$, respectively. The form factors themselves were
again measured with $t=14$ to $16$ fits on $10$ of these lattices. We found
$F=0.80(3)$ and $G_{c}=0.65(8)$, which gives $F/F^{VD}=1.04(4)$ for the pion and
$G_{c}/G_{c}^{VD}=0.86(11)$ for the rho meson. When converted
into an implied monopole mass, these results give $m_{M}/m_{\rho}=1.09(11)$ for
the pion and $m_{M}/m_{\rho}=0.76(13)$ for the rho. The error bars on the
$m_{M}/m_{\rho}$ ratios increased significantly on the $20^{3}\times 30$ lattices
both because a smaller number of configurations were used and because standard
error propagation shows that the relative error in the monopole mass is $\delta
m_{M}/m_{M}=\delta F/[2F(1-F)]$ in terms of the form factor, $F$. The
$(1-F)$ quantity in the denominator increases the error bars at lower momenta.
However, comparing with the $\kappa
=0.152$ results in Tables III and V, we see that both $m_{M}/m_{\rho}$ ratios
have increased. This is what one expects if vector dominance is to be reinstated
at lower $|q^{2}|$.

\section{Summary and Comments}

We have calculated the pion and rho meson form factors on a $16^{3}\times 24$
lattice using the methods of current overlap. We limited our measurements to 
the lowest lattice momentum because of a possible violation of continuum dispersion
seen in the pion local $(E-m)$ measurement. We have extracted a very clean pion
signal which seems to be single exponential even near the time edges. In contrast,
the rho meson electric signals are relatively noisy and seem to be more strongly
affected by time-edge effects. We saw that the combined
$Q(\pm ;q_{z}^{2},t)$ and $Q(0;q_{x,y}^{2},t)$
values had a strong overlap with the $Q(\pm ;q_{x,y}^{2},t)$ results, consistent
with a zero quadrupole form factor. However, because the correct exponential
behavior in the latter quantity at $\kappa=0.154$ and $0.152$ was not
demonstrated in the data, a signal for the quadrupole form factor could not be
isolated. We also saw that the rho magnetic correlation functions,
$K_{y}(+;q_{x},t)$, develop faster in time than either the pion or rho electric
ones, reminiscent of a result in Ref.~\cite{nucleon}. Our error bars on the
magnetic correlation function were encouraging; assuming Eq.(\ref{geqn}), we
extracted a rho meson g-factor and found
$g=2.25(34)$. This value is consistent with the nonrelativistic quark model value
of $2$ and with the recent results of a light-front model\cite{g}.

Three sources of systematic error in our results were
investigated. First, by comparing pion source time separations
of $\Delta T=17$ and $23$, we saw no statistically significant difference in the
time signals for the lowest momentum. We also found that the lattice was long
enough to filter out the correct local $(E-m)$ for both the pion and
rho meson. Second, by using the lattice spin $0$ dispersion relation and
definitions of kinematical factors, we saw that the physical $m_{M}/m_{\rho}$
ratios for the pion and rho meson were decreased by
$3.6\%$ and $2.7\%$, respectively; the rho meson g-factor was increased by
$6.6\%$. In addition, our results on the $20^{3}\times 30$ lattices
at $\kappa=0.152$ gave larger values of $m_{M}/m_{\rho}$ than the smaller lattices,
but with larger error bars. Systematics associated with quenching and scaling have
not been investigated and must wait future studies.

A comparison with the pion and rho results in Ref.~\cite{terry1} has been carried
out. A simple uncorrelated linear extrapolation\cite{bev} of the
Ref.~\cite{terry1} data displayed in Figs.9 and 10 yields a $m_{M}/m_{\rho}$
ratio of $1.08(8)$ for the pion and $0.96(7)$ for the rho. (The CURFIT routine 
from Ref.~\cite{bev} was used for the fit and error bars.) Our
results for these same quantities are $0.97(8)$ and $0.73(10)$. If we combine
errors in an uncorrelated fashion, Ref.~\cite{terry1} finds a rho meson to pion
charge radius of $1.13(12)$, whereas our results give $1.33(21)$. The results in
Ref.~\cite{terry1} were extracted at $|q^{2}|\approx 0.16$ GeV$^{2}$, whereas here
$|q^{2}|\approx 0.47$ GeV$^{2}$. Figs.9 and 10 would seem to indicate a
systematic difference in the $m_{M}/m_{\rho}$ results at finite $ma$.
However, both studies have chirally extrapolated results for the pion which are
consistent with vector dominance and with one another, within large errors. On the
other hand, our extrapolated rho meson $m_{M}/m_{\rho}$ is approximately
$2.7\sigma$ low compared to vector dominance, whereas Ref.~\cite{terry1} is
consistent with the vector dominance prediction. We noted above that our
$20^{3}\times 30$ lattice results hint that the larger $|q^{2}|$ of the present
study may be responsible for the differences seen.

It may help to put our results in the context of measurements of similar
quantities in the baryon case. We are concentrating on the dimensionless
quantity $m_{M}/m_{\rho}$ in the present work, similar to the dipole mass
ratio $m_{D}/m_{N}$ that was studied for the nucleon in
Ref.~\cite{nucleon}. On the same size of lattice as here, it was found in
Ref.~\cite{nucleon} that the $m_{D}/m_{N}$ measurement was about $7\%$ low
compared to experiment after chiral extrapolation. The extrapolated g-factors
there were about $10-20\%$ smaller in magnitude than experiment. It is reasonable
to expect there will be similar sized overall systematic errors on the
$m_{M}/m_{\rho}$ ratios and the rho meson g-factor in our case.
  
Comparison with the experimental results
for the pion form factor\cite{experiment} are suggestive of the
finding of a small violation of n\"aive vector dominance in the spacelike pion
form factor. Experimentally, the the $m_{M}/m_{\rho}$ ratio is $0.96(1)$.
Clearly, the error bars on lattice studies must be significantly
reduced before the results can begin to have phenomenological
impact in this sector.

The simplest explanation for a small violation of n\"aive vector dominance in the
pion form factor involves radially excited states of the rho$^{5}$. Such
physics is contained in the lattice simulation, even in the quenched
approximation. Thus, it is conceivable that the nonzero $ma$ results are indicative
of this violation at our $|q^{2}|$. In any case, there must be additional physics,
outside of n\"aive vector dominance influencing the pion and the rho, resulting in
their different charge radii. Making stronger contact with these experimental
issues will be a challenge for lattice QCD in the future and will require
significantly larger statistics to probe closer to the chiral limit and to
substantially reduce statistical error bars.

\section{Acknowledgments}

This research was supported in part by NSF grant NSF-9401068. W.\ W.\ would like
to thank the theory group at CEBAF and the Department of Physics, University of
Kentucky, for their hospitality. The computations were carried out at the National
Center for Supercomputing Applications and used the CRAY 2, CM5 and Power Challenge
computers. We thank T.\ Draper for allowing us to use his Coulomb gauge fixed
configurations and we acknowledge helpful conversations with K.\ F.\ Liu.

\newpage

\begin{center}
{\bf Footnotes} \\
\end{center}
\begin{enumerate}

\item  New techniques such as in S.\ J.\ Dong and K.\ F.\ Liu, {\it Phys.\
Lett.} {\bf B328} (1994) 130, and Y.\ Kuramashi et.\ al., {\it Phys.\
Rev.\ Lett.} {\bf 72} (1994) 3448, are beginning to change the situation for the
better.

\item We are grateful to T.\ Draper for telling us about this technique.

\item Note that it is always possible to fix a single spatial
point in any n-point correlation function without introducing momentum smearing.

\item Such an assumption is implicit also in the results of
Ref.~\cite{terry1} for the rho meson.

\item See for example Ref.~\cite{last} where two additional vector mesons with
masses of about 1420 and 1770 MeV are sufficient to explain the timelike pion form
factor data in the energy region $1.35 \le \sqrt{s}\le 2.4$ GeV.

\end{enumerate}
\newpage
\begin{table}
\caption{Hadron masses. $\chi^{2}_{d}$ gives the chi-squared
per degree of freedom for the fit.}\label{table1}
\begin{tabular}{cccc}
\multicolumn{1}{c}{Particle}
&\multicolumn{1}{c}{$\kappa=0.154$}
&\multicolumn{1}{c}{$0.152$} &\multicolumn{1}{c}{$0.148$} \\ \hline
pion   &  $0.366(10)$ & $0.479(8)$  &$0.676(6)$\\
$\chi^{2}_{d}$ & $.45$ & $.43$       & $.28$   \\ \\
rho    &  $0.46(2)$  &$ 0.55(1) $ & $0.718(8) $\\
$\chi^{2}_{d}$ & $.23$ & $.19$       & $.05$\\
\end{tabular}
\end{table}
\begin{table}
\protect
\caption{Pion form factors for the lowest spatial momentum.
$F^{VD}$ indicates the vector dominance value for the lattice value of $m_{\rho}$
from Table I. $\chi^{2}_{d}$ gives the chi-squared
per degree of freedom for the fit.}\label{table2}
\begin{tabular}{cccc}
\multicolumn{1}{c}{Quantity}&\multicolumn{1}{c}{$\kappa=0.154$}
&\multicolumn{1}{c}{$\kappa=0.152$}&\multicolumn{1}{c}{$\kappa=0.148$} \\ \hline
$F$ & $0.58(4)$ & $0.65(2)$ &$0.74(3)$ \\
$\frac{F}{F^{VD}_{1}}$ & $0.91(7)$ & $0.94(4)$ &$0.94(3)$ \\
$\chi^{2}_{d}$ & $0.65$ & $0.20$       & $1.02$   \\
\end{tabular}
\end{table}
\begin{table}
\caption{Values of monopole mass divided by rho meson mass, $m_{M}/m_{\rho}$ for
the lowest momentum fit to the pion form factor. $\kappa_{c}$
indicates the correlated chiral extrapolation.}\label{table3}
\begin{tabular}{cccc}
\multicolumn{1}{c}{$\kappa_{c}$}&
\multicolumn{1}{c}{$\kappa=0.154$}
&\multicolumn{1}{c}{$\kappa=0.152$}&\multicolumn{1}{c}{$\kappa=0.148$} \\ \hline
$0.97(8)$ & $ 0.89(6)$ & $0.91(5)$ & $0.89(6)$ \\
\end{tabular}
\end{table}
\begin{table}
\caption{$G_{c}$ and $G_{m}$ are the charge and magnetic form factors of the
rho meson. $G^{VD}$ indicates the vector dominance value for the lattice value of
$m_{\rho}$ from Table I. $\chi^{2}_{d}$ gives the chi-squared
per degree of freedom for the fit.}\label{table4}
\begin{tabular}{cccc}
\multicolumn{1}{c}{Quantity}&\multicolumn{1}{c}{$\kappa=0.154$}
&\multicolumn{1}{c}{$\kappa=0.152$}&\multicolumn{1}{c}{$\kappa=0.148$} \\ \hline
$G_{c}$ & $0.46(6)$ & $0.55(5)$ &$0.68(4)$ \\
$\frac{G_{c}}{G^{VD}}$ & $0.74(9)$ & $0.81(7)$ &$0.87(6)$ \\
$\chi^{2}_{d}$ & $0.43$ & $0.33$       & $0.23$   \\ \\
$G_{m}$ & $0.93(15)$ & $1.07(9)$ & $1.14(7)$ \\
$\chi^{2}_{d}$ & $0.79$ & $0.83$       & $0.79$   \\
\end{tabular}
\end{table}
\begin{table}
\caption{$m_{M}/m_{\rho}$ is the ratio of monopole mass to rho meson mass obtained
from the $G_{c}$ value for the rho meson. The g-factor assumes Eq.(30).}
\label{table5}
\begin{tabular}{ccccc}
\multicolumn{1}{c}{Quantity}&\multicolumn{1}{c}{$\kappa_{c}$}&\multicolumn{1}{c}
{$\kappa=0.154$}
&\multicolumn{1}{c}{$\kappa=0.152$}&\multicolumn{1}{c}{$\kappa=0.148$} \\ \hline
$m_{M}/m_{\rho}$ &$0.73(10)$ & $0.73(8)$ & $0.75(7)$ &$0.77(8)$ \\
g-factor &$2.25(34)$ & $2.03(38)$ & $1.93(21)$ & $1.67(10)$ \\
\end{tabular}
\end{table}

\pagebreak
\begin{center} 
{\bf Figure Captions} 
\end{center}
     
\begin{enumerate}

\item   (a) Graphical representation of the four-point function being calculated.
The time positions of the particle interpolation fields (shown as shaded
circles) are fixed, but the charge and current densities, located on different
flavor lines and symbolized by the \lq\lq X", may be time separated. (b) Example
of a disconnected (sea quark) contribution. The disconnected quark loop is actually
connected to the quarks by an arbitrary number of gluon lines. 

\item  (a) Fourier transform of pion charge overlap function at zero relative
time separation, $Q(\vec{q}\,^{2},0)$, for smeared-to-smeared interpolation
fields ($\kappa=0.154$) as the charge densities are moved together in
time between the sources. ($t=0$ labels the first time slice of the lattice.) (b)
Fourier transform of the same quantity for smeared-to-point interpolation fields.
(The point interpolation field, located at $t=24$, is unsummed.)

\item   Comparison of pion overlap correlation function,
$Q(\vec{q}\,^{2},t)$, versus relative time separation, $t$, for a time separation
$\Delta T=17$ (squares) between the interpolation fields and a time separation of
$\Delta T=23$ (diamonds). Both calculations use smeared-to-smeared quark
propagators with $\kappa=0.154$.

\item   Local $(E-m)$ measurements for the pion overlap function,
$Q(\vec{q}\,^{2},t)$, versus relative time separation, $t$, compared with continuum
dispersion for (a) $|\vec{q}_{min}|=\frac{\pi}{8}$ and (b)
$|\vec{q}|=\sqrt{2}|\vec{q}_{min}|$. Squares are $\kappa=0.154$,
triangles are $\kappa=0.152$, and circles are $\kappa=0.148$ results. Continuum
dispersion is indicated by the solid lines; the associated spin $0$ lattice
dispersion relation is indicated by dotted lines underneath.
 
\item   Correlation functions, $Q(\vec{q}\,^{2},t)$, for the lowest
spatial momenta ($|\vec{q}|=\frac{\pi}{8}$) of the pion versus relative time
separation, $t$. Results are at $\kappa=0.148$ (circles), $\kappa=0.152$
(triangles) and $\kappa=0.154$ (squares). (The kinematical factor
$\frac{(E_{q}+m_{\pi})^{2}}{4E_{q}m_{\pi}}$ has been removed from
$Q(\vec{q}\,^{2},t)$.) The best uncorrelated single exponential fits on the 14-16
time interval are shown.

\item   The electric correlation function $Q(\pm ;q_{z}^{2},t)$ for the rho
meson for the lowest spatial momenta ($|\vec{q}|=\frac{\pi}{8}$) versus relative
time separation, $t$. Results are at
$\kappa=0.148$ (circles), $\kappa=0.152$ (triangles) and
$\kappa=0.154$ (squares). (The kinematical factor
$\frac{(E_{q}+m_{\rho})^{2}}{4E_{q}m_{\rho}}$ has been removed from $Q(\pm
;q_{z}^{2},t)$.) The best single exponential fits on the 14-16 time interval are
shown for $\kappa=0.148$ and $0.152$; the best fit for the 13-15 interval is
shown for $\kappa=0.154$.

\item   The magnetic correlation function $K_{y}(+ ;q_{x},t)$ for the rho meson
for the lowest spatial momentum ($|\vec{q}|=\frac{\pi}{8}$) versus relative
time separation, $t$. Results are at $\kappa=0.148$ (circles), $\kappa=0.152$
(triangles) and $\kappa=0.154$ (squares). (The
factor $iq_{x}\frac{(E_{q}+m_{\rho})}{4E_{q}m_{\rho}}$ has been removed from
$K_{y}(+ ;q_{x},t)$.) The best single exponential fits on the 7-9 time interval
are shown.

\item   The quadrupole correlation function $Q_{\rho}(\pm ;q_{x,y}^{2},t)$ for the
rho meson for the lowest spatial momentum ($|\vec{q}|=\frac{\pi}{8}$) versus 
relative time separation, $t$. (The kinematical factor
$\frac{(E_{q}+m_{\rho})^{2}}{4E_{q}m_{\rho}}$ has been removed.) Results are at
$\kappa=0.148$ (circles), $\kappa=0.152$ (triangles) and
$\kappa=0.154$ (squares). No fits of these functions were made.

\item   Monopole mass divided by the rho meson mass, $m_{M}/m_{\rho}$, for the
pion as a function of dimensionless quark mass ($ma$) as determined from the lowest
momentum measurements (squares). The darkened square gives the chiral
extrapolation. The triangles give results extracted from Table 1 of
Ref.~\cite{terry1}.

\item   Monopole mass divided by the rho meson mass, $m_{M}/m_{\rho}$, for the rho
meson as a function of dimensionless quark mass ($ma$) as determined from the
lowest momentum measurements (squares) assuming the quadrupole form factor is
zero. The darkened square gives the chiral extrapolation. Triangles are the
results extracted from Table I of Ref.~\cite{terry1}.

\item   g-factor for the rho as a function of dimensionless quark mass
($ma$) assuming Eq.(\ref{geqn}). Darkened square gives the chiral extrapolation. 

\end{enumerate}
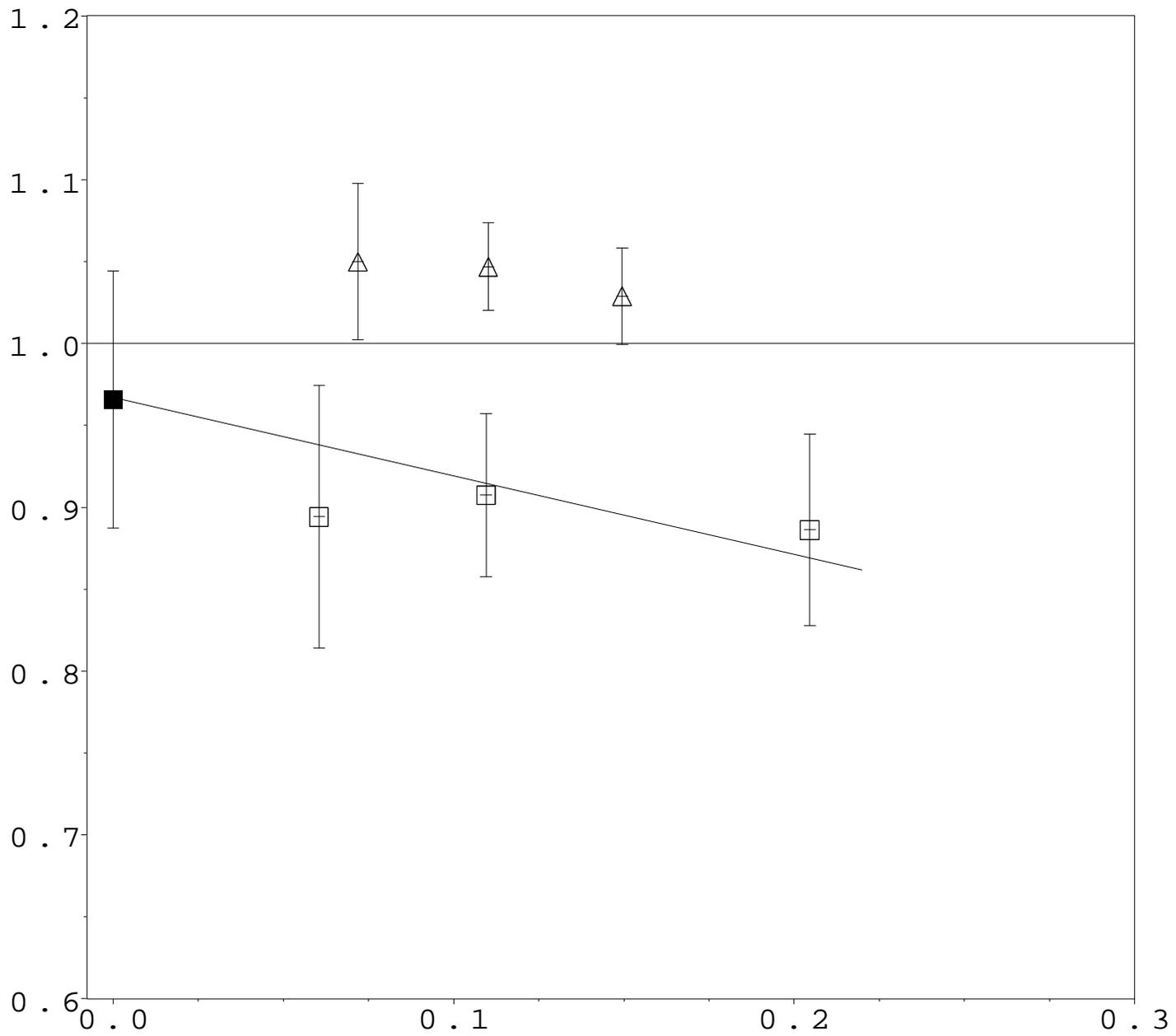
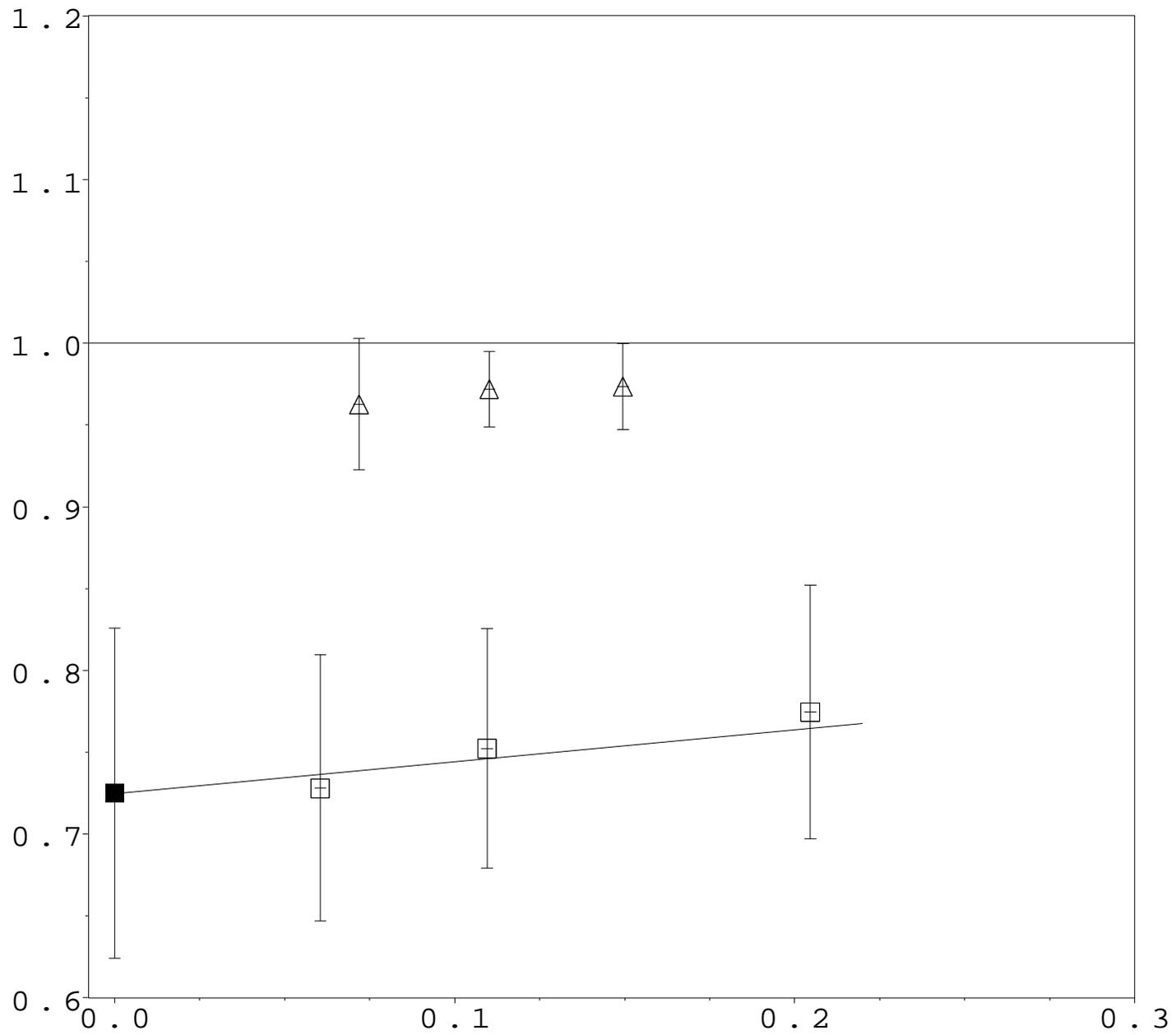
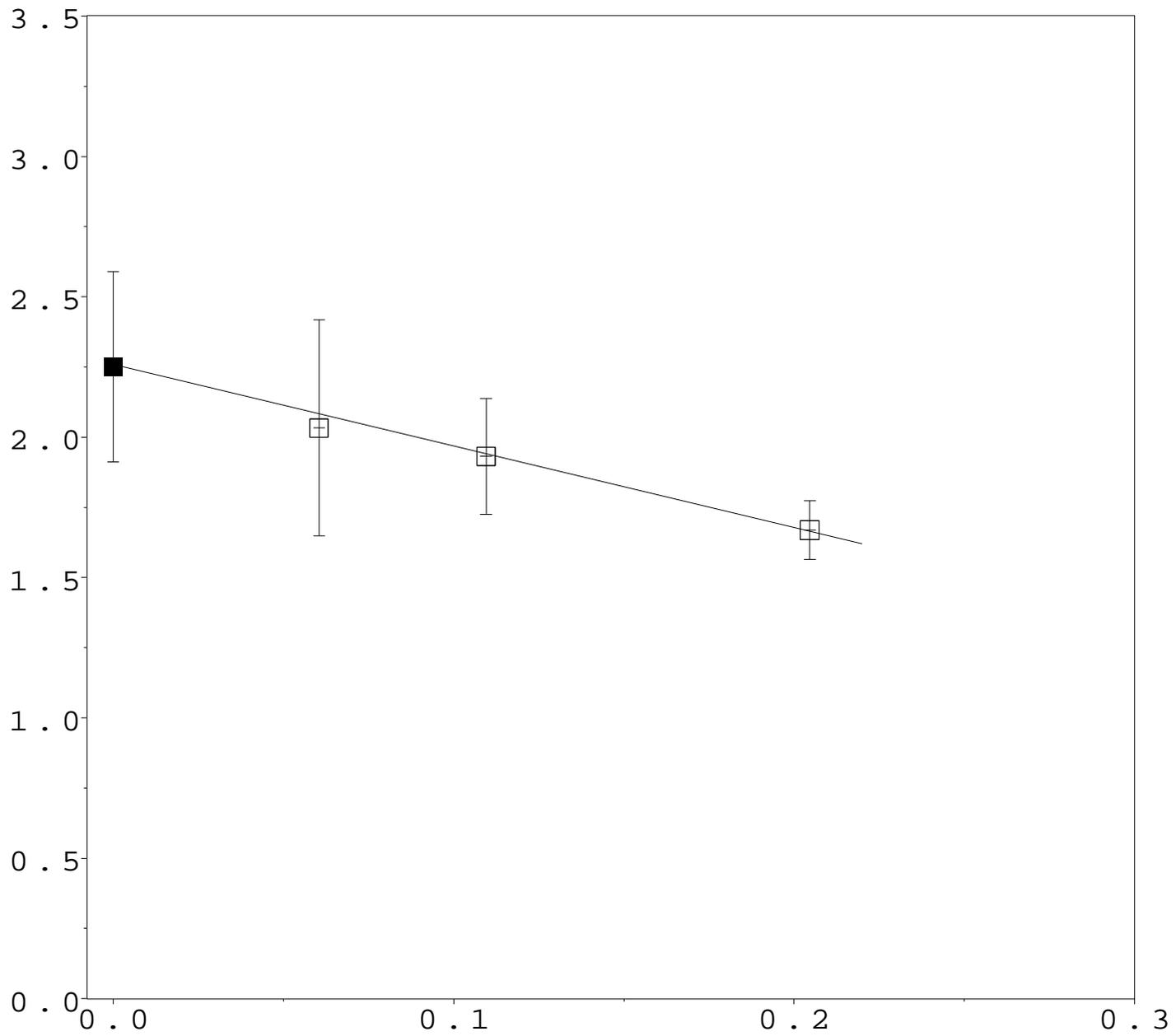
\end{document}